\documentstyle[12pt]{article}
 \def\nh{\noindent\hangindent=0.45truecm\hangafter=1}
 \def\sqr#1#2{{\vcenter{\hrule height.#2pt \hbox{\vrule width.#2pt
 height#2pt \kern#1pt \vrule width.#2pt} \hrule height.#2pt}}}

\newtheorem{theo}{\sc Theorem}
\newtheorem{lemm}{\sc Lemma}

\newcommand{\tod}{\stackrel{d}{\longrightarrow}}


\def\be{\begin{equation}}
  \def\ee{\end{equation}}
  \def\nn{\nonumber}
  \def\bea{\begin{eqnarray}}
  \def\eea{\end{eqnarray}}
  
\begin{document}

\begin{center}
 {\bf \large Empirical likelihood for linear models with spatial errors}
\end{center}

\bigskip

\begin{center}
Yongsong Qin
\footnote{Corresponding author. Tel: +86-773-5851556

\ \ \  Email address: ysqin@gxnu.edu.cn (Y. Qin)}
\bigskip

{\small  Department of Statistics, Guangxi Normal University\\
Guilin, Guangxi 541004, China}

\end{center}
\bigskip

 \noindent
{\bf Abstract }  For linear models with spatial errors, the empirical likelihood ratio statistics are constructed for the parameters of the models. It is shown that the limiting distributions of the empirical likelihood ratio statistics are chi-squared distributions, which are used to construct confidence regions for the parameters of the models.



\noindent {\it Short title:} Empirical likelihood for linear models with spatial errors

\newpage

\setcounter{section}{1}

 \noindent
{\bf 1. Introduction}
\bigskip

The linear regression models are the most important statistical models for explaining the relationship between response and explanatory variables.
Whenever the variables in a linear regression model refer
to attributes of a particular location (height of a plant,
population of a country, position in a social network, etc.),
one often allows for correlation among the errors (disturbances)
by assuming that the errors follow a spatial autoregressive correlation (e.g. Dow et al., 1982; Ord, 1975; Kr\"{a}mer and Donninger, 1987). Then we have
the following linear regression model with spatial autoregressive errors:
\be
Y_n=X_n\beta+u_{(n)}, u_{(n)}=\rho W_n u_{(n)}+\epsilon_{(n)}, \label{model1}
\ee
where $n$ is the number of spatial units,  $\beta$ is the $k\times 1$ vector of regression parameters,  $X_n=(x_1, x_2, \cdots, x_n)^{\tau}$ is the non-random $n\times k$ matrix of observations on the independent variable, $Y_n=(y_1, y_2, \cdots, y_n)^{\tau}$ is an $n\times 1$ vector of observations on the dependent variable, $u_{(n)}$ is an $n\times 1$ vector of errors (disturbances), $\rho$ is the scalar autoregressive parameter with $|\rho|<1$, $W_n$ is an $n\times n$ spatial weighting matrix of constants, $\epsilon_{(n)}$ is an $n\times 1$ vector of innovations which satisfies
\[
E\epsilon_{(n)}=0, Var(\epsilon_{(n)})=\sigma^2I_n.
\]
Model (\ref{model1}) is also called spatial error model (SEM).  The development in testing and estimation of SEM models has been summarized in Anselin (1988), Cliff and Ord (1973), Ord (1975), Kr\"{a}mer and Donninger (1987) and Helejian and Prucha (1999), among others.

There are two competing estimation approaches for the corresponding parameters. One is the maximum likelihood (ML) method (e.g. Anselin, 1988). The other is the computationally more efficient method, the generalized
method of moment (GMM) approach by Kelejian and Prucha (1999). The asymptotic properties of the maximum likelihood estimator (MLE) and the GMM estimator for the SEM model are investigated by Anselin (1988) and Kelejian and Prucha (1999), respectively. However, it may not be easy to use these normal approximation results to construct confidence region for the parameters in the SEM model as the asymptotic covariance in the asymptotic distribution is unknown. More importantly, the accuracy of the normal approximation based confidence region of the parameters in the model  may be affected by estimating the asymptotic covariance. In this article, we propose to use the empirical likelihood (EL) method introduced by Owen (1988, 1990) to construct confidence region for the parameters in the SEM model.
The shape and orientation of the EL confidence region are determined by data and the confidence region is obtained without covariance estimation. These features of the EL confidence region are the major motivations for our current proposal. Owen (1991) has used the EL method to construct confidence regions for the vector of regression parameters in a linear model with independent errors. A comprehensive review on EL for regressions can be found in Chen and Keilegom (2009). More references on EL methods can be found in Owen (2001), Qin and Lawless (1994), Chen and Qin (1993),  Zhong and Rao (2000) and   Wu (2004), among others.

The idea in using the EL method for the SEM is to introduce a martingale sequence to transform the linear-quadratic form of the estimating equations (e.g. (\ref{2.1})-(\ref{2.3})) for the SEM into a linear form.  It is interesting to note that the estimation equations for other spatial models may have the linear-quadratic forms. Therefore this approach of transformation also opens a way to use EL methods to more general spatial models.

The article is organized as follows.  Section 2 presents the
main results. Results from a simulation study are reported in Section 3. All the
technical details are  presented in Section 4.

\bigskip
\setcounter{section}{2}

 \noindent
{\bf 2. Main Results}
\bigskip

We continue with model (\ref{model1}). Let $A_n(\rho)=I_n-\rho W_n$ and suppose that $A_n(\rho)$ is nonsingular. Then
\[
Y_n=X_n\beta+A^{-1}_n(\rho)\epsilon_{(n)}.
\]
At this moment, suppose that $\epsilon_{(n)}$ is normally distributed, which is  used to derive the EL statistic only and not employed in our main results. Then the log-likelihood function based on the response vector $Y_n$ is
\[
L=-{n\over 2}\log (2\pi)-{n\over 2}\log\sigma^2 +\log |A_n(\rho)|-{1\over 2\sigma^2}\epsilon^{\tau}_{(n)}\epsilon_{(n)},
\]
where $\epsilon_{(n)}=A_n(\rho)(Y_n-X_n\beta)$. Let $G_n=W_nA^{-1}_n(\rho)$ and $\tilde{G}_n={1\over 2}(G_n+G^{\tau}_n)$.
It can be shown that (e.g.  Anselin, 1988, pp. 74-75)
\[
\partial L/\partial \beta={1\over \sigma^2}X^{\tau}_nA_n(\rho)\epsilon_{(n)},
\]
\bea
\partial L/\partial \rho &=& {1\over \sigma^2}\{\epsilon^{\tau}_{(n)}W_nA^{-1}_n(\rho)\epsilon_{(n)}-\sigma^2 tr(W_nA^{-1}_n(\rho))\}\nn\\
&=& {1\over \sigma^2}\{\epsilon^{\tau}_{(n)}\tilde{G}_n\epsilon_{(n)}-\sigma^2 tr(\tilde{G}_n)\},\nn
\eea
\[
\partial L/\partial \sigma^2 = {1\over 2\sigma^4}(\epsilon^{\tau}_{(n)}\epsilon_{(n)}-n\sigma^2).
\]
Letting above derivatives be $0$, we obtain the following estimating equations:
\be
X^{\tau}_nA_n(\rho)\epsilon_{(n)}=0, \label{2.1}
\ee
\bea
\epsilon^{\tau}_{(n)}\tilde{G}_n\epsilon_{(n)}-\sigma^2 tr(\tilde{G}_n)=0,
\label{2.2}
\eea
\be
\epsilon_{(n)}^{\tau}\epsilon_{(n)}-n\sigma^2=0.
\label{2.3}
\ee
We use  ${g}_{ij}$,  $\tilde{g}_{ij}$ and $b_i$ to denote the $(i, j)$ element of the matrix ${G}_n$, the $(i, j)$ element of the matrix $\tilde{G}_n$ and the $i$-th column of  the matrix $X^{\tau}_nA_n(\rho)$, respectively,
and adapt the convention that any sum with an upper index of less than one is zero. To deal with the quadratic form in (\ref{2.2}), we follow Kelejian and Prucha (2001) to introduce a martingale difference array. Define the $\sigma$-fields: ${\mathcal{F}}_{0}=\{ {\emptyset}, \Omega\}, {\mathcal{F}}_{i}=\sigma(\epsilon_1, \epsilon_2, \cdots, \epsilon_i), 1\leq i\leq n$. Let
\be
\tilde{Y}_{in}=\tilde{g}_{ii}(\epsilon^2_i-\sigma^2)+2\epsilon_i\sum^{i-1}_{j=1}\tilde{g}_{ij}\epsilon_j.\label{4.1.1}
\ee
Then $ {\mathcal{F}}_{i-1}  \subseteq {\mathcal{F}}_{i}, \tilde{Y}_{in}$ is ${\mathcal{F}}_{i}-$ measurable and $E(\tilde{Y}_{in}|{\mathcal{F}}_{i-1})=0$. Thus $\{\tilde{Y}_{in}, {\mathcal{F}}_{i}, 1\leq i\leq n\}$ form a martingale difference array and
\be
\epsilon^{\tau}_{(n)}\tilde{G}_n\epsilon_{(n)}-\sigma^2 tr(\tilde{G}_n)=\sum^n_{i=1}\tilde{Y}_{in}. \label{4.1.2}
\ee

Based on (\ref{2.1}) to (\ref{4.1.2}), we propose the following EL ratio statistic for $\theta\hat{=}(\beta^{\tau}, \rho, \sigma^2)^{\tau}\in R^{k+2}$:
\[
L_n(\theta)=\sup_{p_i, 1\leq i\leq n}\prod^n_{i=1}(np_i),
\]
where $\{p_i\}$ satisfy
\bea
&& p_i\geq 0, 1\leq i\leq n, \sum^n_{i=1}p_i=1, \nn\\
&&  \sum^n_{i=1}p_i b_i\epsilon_i=0, \nn\\
&& \sum^n_{i=1}p_i \bigg \{\tilde{g}_{ii}(\epsilon^2_i-\sigma^2)+2\epsilon_i\sum^{i-1}_{j=1}\tilde{g}_{ij}\epsilon_j \bigg \}=0,\nn\\
&& \sum^n_{i=1}p_i(\epsilon^2_i-\sigma^2)=0,\nn
\eea

Let
\[
\omega_i(\theta)=\left ( \begin{array}{c}  b_i\epsilon_i\\
\tilde{g}_{ii}(\epsilon^2_i-\sigma^2)+2\epsilon_i\sum^{i-1}_{j=1}\tilde{g}_{ij}\epsilon_j\\
\epsilon^2_i-\sigma^2\\
\end{array} \right )_{(k+2)\times 1},
\]
where $\epsilon_i$ is the $i$-th component of $\epsilon_{(n)}=A_n(\rho)(Y_n-X_n\beta)$.
Following Owen (1990), one can show that
\be
\ell_n(\theta)\hat{=}-2\log L_n(\theta)=2\sum^n_{i=1}\log \{1+\lambda^{\tau}(\theta)\omega_i(\theta)\},  \label{8}
\ee
where $\lambda(\theta)\in R^{k+2}$ is the solution of the following equation:
\be
{1\over n}\sum^n_{i=1}{\omega_i(\theta)\over 1+\lambda^{\tau}(\theta)\omega_i(\theta)}=0.  \label{9}
\ee

Let $\mu_j=E\epsilon^j_1, j=3, 4$. Use $Vec(diag A)$ to denote the vector formed by the diagonal elements of a matrix $A$ and $||a||$ to denote the $L_2$-norm of a vector $a$.  Furthermore, Let  ${\mathbf{1}_n}$ present the $n$-dimensional (column) vector with $1$ as its components. To obtain the asymptotical distribution of $\ell_n(\theta)$, we need
following assumptions.

A1.\  $\{\epsilon_{i}, 1\leq i\leq n\}$ are independent and identically distributed random variables with mean $0$, variance $\sigma^2>0$ and
$E|\epsilon_{1}|^{4+\eta_1}<\infty$ for some $\eta_1>0$.

A2.\  \ Let $W_n$, $A_n^{-1}(\rho)$ and $\{x_i\}$  be as described above. They satisfy the
following conditions:

(i) The row and column sums of $W_n$ and  $A_n^{-1}(\rho)$  are uniformly bounded in absolute value;

(ii) $\{x_i\}$ are uniformly bounded.

A3. There is a constants $c_j>0, j=1, 2$, such that  $0<c_1\leq \lambda_{min}\left ( n^{-1}\Sigma_{k+2} \right )\leq \lambda_{max}\left ( n^{-1}\Sigma_{k+2} \right )\leq c_2<\infty$, where $\lambda_{min}(A)$ and $\lambda_{max}(A)$ denote the  minimum and maximum eigenvalues of a matrix
$A$, respectively,
 \be
\Sigma_{k+2}=\Sigma^{\tau}_{k+2}=Cov\left \{\sum^n_{i=1}\omega_i(\theta) \right \}=\left ( \begin{array}{ccc} \Sigma_{11} \  \Sigma_{12} \   \Sigma_{13}\\
\Sigma_{21}\  \Sigma_{22}  \  \Sigma_{23}\\
\Sigma_{31} \  \Sigma_{32}   \  \Sigma_{33} \end{array}
\right ),  \label{Sigma}
\ee
\[
\Sigma_{11}=\sigma^2X^{\tau}_nA_n(\rho)A^{\tau}_n(\rho)X_n, \Sigma_{12}=\mu_3X^{\tau}_nA_n(\rho)Vec(diag \tilde{G}_n),
\]
\[
 \Sigma_{13}=\mu_3X^{\tau}_nA_n(\rho) {\mathbf{1}}_n,
\Sigma_{22} = 2\sigma^4tr(\tilde{G_n}^2)+(\mu_4-3\sigma^4)||Vec(diag\tilde{G_n})||^2,
\]
\[
\Sigma_{23}=(\mu_4-\sigma^4)tr(\tilde{G_n}), \Sigma_{33}=n(\mu_4-\sigma^4).
\]

{\bf Remark 1. } Conditions A1 to A3 are common assumptions for SAR models. For example, A1 and A2 are used in Assumptions 1, 4, 5 and 6 in Lee (2004), the analog  of $0<c_1\leq \lambda_{min}\left ( n^{-1}\Sigma_{k+2} \right )$ (e.g. $n^{-1}\sigma^2_{\widetilde{Q}}\geq c$ for some constant $c>0$ in Lemma \ref{lemm4.2} in this article) is employed in the assumption of Theorem 1 in Kelejian and Prucha (2001). From Conditions A1 and A2, one can see that $\lambda_{max}\left ( n^{-1}\Sigma_{k+2} \right )\leq c_2<\infty$. For the sake of argument, we list this consequence of  A1 and A2 as a condition here.

We now state the main results.

\begin{theo}\label{theo2.1}
Suppose that Assumptions (A1) to (A3) are satisfied. Then under model (\ref{model1}), as $n\to \infty$,
\[
\ell_n ({\theta} )\tod \chi^2_{k+2},
\]
where $\chi^2_{k+2}$ is a chi-squared distributed random variable with
$k+2$ degrees of freedom.
\end{theo}

Let $z_{\alpha}(k+2)$ satisfy  $P(\chi^2_{k+2}\leq z_{\alpha}(k+2))=\alpha$ for $0<\alpha<1$. It follows from Theorem
\ref{theo2.1} that an EL based confidence region for $\theta$ with
asymptotically correct coverage probability $\alpha$ can be
constructed as
\[
\{ \theta: \ell_n(\theta)\leq z_{\alpha}(k+2) \}.
\]

\bigskip
\setcounter{section}{3}
 \noindent
{\bf 3. Simulations}
\bigskip

According to Anselin (1988), when the error term $\epsilon_{(n)}$  is normal distributed, the likelihood ratio (LR)
$LR(\theta_0)=2(L(\hat\theta)-L(\theta_0))$ is asymptotically distributed as $\chi^2_{k+2}$ under the null hypothesis: $\theta=\theta_0$, where $L$ is the corresponding log-likelihood and $\hat\theta$ is the maximum likelihood estimator.
It follows that the LR based confidence region for $\theta$ with
asymptotically correct coverage probability $\alpha$ can be
constructed as
\[
\{ \theta: LR(\theta)\leq z_{\alpha}(k+2) \}.
\]
We note that the LR method requires to know the form of the distribution of the population in study, while the EL method does not. This fact implies that the EL method performs better than the LR method theoretically when the population distribution is not normal. Our following simulation results do confirm this conclusion. 


We conducted a small simulation study to compare the finite sample performances of the confidence regions based on EL and LR methods with confidence level $\alpha=0.95$, and report the proportion of  $LR(\theta_0)\le z_{0.95}(k+2)$ and $\ell_n (\theta_0) \le z_{0.95}(k+2)$ respectively in our $2,000$ simulations, where $\theta_0$ is the true value of $\theta$. The results of simulations are reported in tables 1 to 3.

In the simulations, we used the model: $Y_n=X_n\beta+u_{(n)}, u_{(n)}=\rho W_n u_{(n)}+\epsilon_{(n)}$ with $X_n=(x_1, x_2, \cdots, x_n)^{\tau}, x_i={i\over n+1}, 1\leq i\leq n$,  $\beta=3.5$,  $\rho$ were taken as $-0.85$, $-0.15$£¬ $0.15$ and $0.85$, respectively,  and $\epsilon_i's$ were taken from $ N(0,1)$, $t(5)$ and $\chi^2_{4}-4$, respectively.

For the contiguity weight matrix $W_n=(W_{ij})$, we took $W_{ij}=1$ if  spatial units $i$ and $j$ are neighbours by queen contiguity rule (namely, they share common border or vertex), $W_{ij}=0$ otherwise (Anselin, 1988, P.18). We first considered three ideal cases of spatial units: $n=m\times m$ regular grid with $m=7, 10, 13$, denoting $W_n$ as $grid_{49}, grid_{100} $ and $ grid_{169}$, respectively. Secondly, we used the weight matrix $W_{49}$ related to $49$ contiguous planning neighborhoods in Columbus, Ohio, U.S., which appeared in Anselin(1988, P. 187). Thirdly, $W_n=I_5\bigotimes W_{49}$ was considered, where $\bigotimes$ is kronecker product. This corresponds to the pooling of
five separate districts with similar neighboring structures in each district. Finally, weight matrix $W_{345}$ was included in the simulations, which is related to $345$ major cities in China.

  A transformation is often used in applications to convert the matrix $W_n$  to the unity of row-sums. We used the standardized version of $W_n$ in our simulations, namely $W_{ij}$ was replaced by $W_{ij}/\sum_{j=1}^nW_{ij}$.

 Simulation results show that the confidence regions based on LR behave  well  with  coverage probabilities very close to the nominal level 0.95 when the error term $\epsilon_i$ is normally distributed, but not well in other cases. The coverage probabilities of the confidence regions based on LR fall to the range [0.8045,0.8560] for $t$ distribution and [0.8295, 0.8615] for $\chi^2$ distribution, which are far from the nominal level 0.95.

 We can see, from tables 1 to 3 , the confidence regions based on EL method converge to the nominal level $0.95$ as the number of spatial units $n$ is large enough, whether the error term $\epsilon_i$ is normally distributed or not.
Our simulation results recommend EL method when we can not confirm the normal distribution of the  error term.

\bigskip

{\bf Tables 1-3 are about here.}

\bigskip
\setcounter{section}{4}
 \noindent
{\bf 4. Proofs}
\bigskip

In the proof of the main results, we need to use Theorem $1$ in Kelejian and Prucha (2001). We now state this result. Let
\[
\widetilde{Q}_n=\sum^n_{i=1}\sum^n_{j=1}a_{nij}\epsilon_{ni}\epsilon_{nj}+\sum^n_{i=1}b_{ni}\epsilon_{ni},
\]
where $\epsilon_{ni}$ are real valued random variables, and the $a_{nij}$ and $b_{ni}$ denote the real valued coefficients of the linear-quadratic form.
We need the following assumptions in Lemma \ref{lemm4.2}.

(C1)  $\{\epsilon_{ni}, 1\leq i\leq n\}$ are independent random variables with mean $0$ and $\sup_{1\leq i\leq n, n\geq 1}E|\epsilon_{ni}|^{4+\eta_1}<\infty$ for some
$\eta_1>0$;

(C2)  For all $1\leq i, j\leq n, n\geq 1, a_{nij}=a_{nji}$, $\sup_{1\leq j\leq n, n\geq 1} \sum^n_{i=1}|a_{nij}|<\infty$, and $\sup_{n\geq 1}n^{-1}\sum^n_{i=1}|b_{ni}|^{2+\eta_2}<\infty$ for some $\eta_2>0$.

Given the above assumptions (C1) and (C2), the mean and variance of $\tilde{Q}_n$ are given  as (e.g. Kelejian and Prucha, 2001)
\[
\mu_{\widetilde{Q}}=\sum^n_{i=1}a_{nii}\sigma^2_{ni},
\]
\bea
\sigma^2_{\widetilde{Q}} &=& 2\sum^n_{i=1}\sum^{n}_{j=1}a^2_{nij}\sigma^2_{ni}\sigma^2_{nj}+\sum^n_{i=1}b^2_{ni}\sigma^2_{ni}\nn\\
&&  +\sum^n_{i=1}\{ a^2_{nii}(\mu^{(4)}_{ni}-3\sigma^4_{ni})+2b_{ni}a_{nii}\mu^{(3)}_{ni}  \}, \label{variance}
\eea
with $\sigma^2_{ni}=E(\epsilon_{ni}^2)$ and $\mu^{(s)}_{ni}=E(\epsilon_{ni}^s)$ for $s=3, 4$.

\begin{lemm}\label{lemm4.2}
Suppose that Assumptions C1 and C2 hold true and $n^{-1}\sigma^2_{\widetilde{Q}}\geq c$ for some constant $c>0$. Then
\[
{\widetilde{Q}_n-\mu_{\widetilde{Q}}\over \sigma_{\widetilde{Q}} } \tod N(0, 1).
\]
\end{lemm}

{\bf Proof. } See Theorem $1$ and Remark $12$ in Kelejian and Prucha (2001).

\begin{lemm}\label{lemm4.3}
Let $\eta_1, \eta_2,\cdots, \eta_n$  be a sequence of stationary random variables,
 with $E|\eta_1|^s<\infty$ for some constants $s>0$ and $C>0$.  Then
\[
\max_{1\leq i \leq n }|\eta_i|=o(n^{1/s}), \ \   a.s.
\]
\end{lemm}

{ \bf Proof. } It is straightforward.

\begin{lemm}\label{lemm4.4}
Suppose that Assumptions (A1) to (A3) are satisfied. Then as $n\to \infty$,
\bea
Z_n=\max_{1\leq i \leq n}||\omega_i(\theta)||=o_p(n^{1/2})\ \ a.s., \label{21}
\eea
\bea
\Sigma_{k+2}^{-1/2}{\sum_{i=1}^n \omega_i(\theta)}\tod N(0, I_{k+2}),\label{22}
\eea
\bea
n^{-1}\sum_{i=1}^n \omega_i(\theta)\omega_i^\tau(\theta)=n^{-1}\Sigma_{k+2}+o_p(1),\label{23}
\eea
\bea
\sum_{i=1}^n ||\omega_i(\theta)||^3=O_p(n),\label{24}
\eea
where $\Sigma_{k+2}$ is given in (\ref{Sigma}).
\end{lemm}

{\bf Proof. } Note that
\bea
Z_n &\leq &\max_{1\leq i \leq n}||b_i\epsilon_i||+\max_{1\leq i \leq n}\left |\tilde{g}_{ii}(\epsilon^2_i-\sigma^2)+2\epsilon_i\sum^{i-1}_{j=1}\tilde{g}_{ij}\epsilon_j \right |\nn\\
&& +
\max_{1\leq i \leq n}|\epsilon^2_i-\sigma^2|\nn\\
&\leq& \max_{1\leq i \leq n}||b_i\epsilon_i||+\max_{1\leq i \leq n}|\tilde{g}_{ii}(\epsilon^2_i-\sigma^2)|+\max_{1\leq i \leq n}
\left |2\epsilon_i\sum^{i-1}_{j=1}\tilde{g}_{ij}\epsilon_j \right |\nn\\
&&+
\max_{1\leq i \leq n}|\epsilon^2_i-\sigma^2|.\nn
\eea
By Conditions A1 and A2 and Lemma \ref{lemm4.3}, we have
\[
\max_{1\leq i \leq n}||b_i\epsilon_i||=\max_{1\leq i \leq n}||b_i||o_p(n^{1/4})=o_p(n^{1/4}),
\]
\[
\max_{1\leq i \leq n}|\tilde{g}_{ii}(\epsilon^2_i-\sigma^2)|=\max_{1\leq i \leq n}|\tilde{g}_{ii}|o_p(n^{1/2})=o_p(n^{1/2}),
\]
\[
\max_{1\leq i \leq n}\left |\epsilon_i\sum^{i-1}_{j=1}\tilde{g}_{ij}\epsilon_j \right |=(\max_{1\leq i \leq n}|\epsilon_i|)^2\cdot
\max_{1\leq i \leq n}\left |\sum^{i-1}_{j=1}\tilde{g}_{ij} \right |=o_p(n^{1/2}),
\]
\[
\max_{1\leq i \leq n}|\epsilon^2_i-\sigma^2|=o_p(n^{1/2}),
\]
Thus $Z_n=o_p(n^{1/2})$. (\ref{21}) is proved.

For any given ${l}=(l^\tau_1,l_2,l_3)^\tau\in R^{k+2}$ with $||{l}||=1$, where $l_1\in R^{k}, l_2, l_3\in R$.
Then
\bea
l^\tau\omega_i(\theta)&&=l_1^\tau b_i\epsilon_i+l_2\{\tilde{g}_{ii}(\epsilon^2_i-\sigma^2)+2\epsilon_i
\sum^{i-1}_{j=1}\tilde{g}_{ij}\epsilon_j\}+l_3(\epsilon^2_i-\sigma^2)\nn\\
&&=(l_2\tilde{g}_{ii}+l_3)(\epsilon^2_i-\sigma^2)+2\epsilon_i\sum^{i-1}_{j=1}l_2\tilde{g}_{ij}\epsilon_j
+l_1^\tau b_i\epsilon_i.\nn
\eea
Thus
\bea
\sum_{i=1}^n l^\tau\omega_i(\theta)=\sum_{i=1}^n (l_2\tilde{g}_{ii}+l_3)(\epsilon^2_i-\sigma^2)+2\sum_{i=1}^n \sum^{i-1}_{j=1}l_2\tilde{g}_{ij}\epsilon_i\epsilon_j
+\sum_{i=1}^n l_1^\tau b_i\epsilon_i.\nn
\eea
Let
\[
Q_n= \sum^n_{i=1}\sum^{n}_{j=1}u_{ij}\epsilon_i\epsilon_j+\sum^n_{i=1}v_{i}\epsilon_i ,
\]
where
\[
u_{ii}=l_2\tilde{g}_{ii}+l_3, u_{ij}=l_2\tilde{g}_{ij}(i\neq j), v_i=l_1^\tau b_i.
\]
Then
\[
Q_n=\sum_{i=1}^n l^\tau\omega_i(\theta)=\sum_{i=1}^n \{u_{ii}(\epsilon^2_i-\sigma^2)+\sum^{i-1}_{j=1}u_{ij}\epsilon_i\epsilon_j+v_{i}\epsilon_i\}.
\]
 To obtain the asymptotic distribution of $Q_n$, we  need to check Condition C2.
From Condition A2(i), it can be shown that
\bea
\sum^n_{i=1}|u_{ij}| &\leq & |l_2|\sum^n_{i=1}|\tilde{g}_{ij}|+|l_3|\leq C. \label{4.1}
\eea
Further,
\bea
n^{-1}\sum^n_{i=1}|v_{i}|^3=n^{-1}\sum^n_{i=1}|l_1^{\tau}b_i|^3
\leq
 C \max_{1\leq i\leq n}||x_i||^3\max_{1\leq i\leq n}(\sum_{k=1}^n |{a}_{ik}|)^3\leq C, \label{4.2}
\eea
where ${a}_{ik}$ is the $(i, k)$-element of $A_n(\rho)$.
From (\ref{4.1}) and (\ref{4.2}), it follows that $n^{-1}\sum^n_{i=1}|v_{i}|^3\leq C$.
Therefore, Condition C2 is satisfied.

We now derive the variance of $Q_n$. Let $e_i$ be the unit vector in the $i$-th coordinate direction.  It can be shown that
\bea
\sum^n_{i=1}\sum^{n}_{j=1}u^2_{ij} &=& \sum^n_{i=1}\{(l_2\tilde{g}_{ii}+l_3)^2+\sum_{i\neq j}(l_2\tilde{g}_{ij})^2\}\nn\\
&=& \sum^n_{i=1}\{(l_2\tilde{g}_{ii})^2+2l_2l_3\tilde{g}_{ii}+l_3^2+\sum_{i\neq j}(l_2\tilde{g}_{ij})^2\}\nn\\
&=& 2l_2l_3\sum^n_{i=1}\tilde{g}_{ii}+nl_3^2+\sum^n_{i=1}\sum^{n}_{j=1}(l_2\tilde{g}_{ij})^2\nn\\
&=& 2l_2l_3tr(\tilde{G_n})+nl_3^2+l_2^2tr(\tilde{G_n}^2),\nn
\eea
\bea
\sum^n_{i=1}u^2_{ii} &=& \sum^n_{i=1}(l_2\tilde{g}_{ii}+l_3)^2\nn\\
&=& l_2^2\sum^n_{i=1}\tilde{g}_{ii}^2+2l_2l_3tr(\tilde{G_n})+nl_3^2\nn\\
&=& l_2^2||Vec(diag\tilde{G_n})||^2+2l_2l_3tr(\tilde{G_n})+nl_3^2,\nn
\eea
\bea
 \sum^n_{i=1}v^2_{i} &=&\sum^n_{i=1}(l_1^{\tau}b_i)^2=l_1^{\tau}\left ( \sum^n_{i=1}b_i b^{\tau}_i\right )l_1\nn\\
&=& l_1^{\tau}\left ( \sum^n_{i=1}X^{\tau}_nA_n(\rho)e_i e^{\tau}_iA^{\tau}_n(\rho)X_n\right )l_1\nn\\
&=& l_1^{\tau}X^{\tau}_nA_n(\rho)\left ( \sum^n_{i=1}e_i e^{\tau}_i\right )A^{\tau}_n(\rho)X_nl_1\nn\\
&=& l_1^{\tau}X^{\tau}_nA_n(\rho)A^{\tau}_n(\rho)X_nl_1,\nn
\eea
and that
\bea
\sum^n_{i=1}u_{ii}v_{i} &=& \sum^n_{i=1}(l_2\tilde{g}_{ii}+l_3)l_1^{\tau}b_i\nn\\
&=&   l_1^{\tau} X^{\tau}_nA_n(\rho)Vec(diag \tilde{G}_n)l_2+l_1^{\tau}X^{\tau}_nA_n(\rho) {\mathbf{1}}_n l_3,\nn
\eea
where  ${\mathbf{1}_n}$ is the $n$-dimensional vector with $1$ as its components.
It follows from (\ref{variance}) that the variance of $Q_n$ is
\bea
\sigma^2_Q &=& 2\sum^n_{i=1}\sum^{n}_{j=1}u^2_{ij}\sigma^4+\sum^n_{i=1}v^2_{i}\sigma^2
 +\sum^n_{i=1}\{ u^2_{ii}(\mu_4-3\sigma^4)+2u_{ii}v_{i}\mu_3  \}\nn\\
 &=& 2\sigma^4\{l_2^2tr(\tilde{G_n}^2)+2l_2l_3tr(\tilde{G_n})+nl_3^2\}\nn\\
 &&+\sigma^2l_1^{\tau}X^{\tau}_nA_n(\rho)A^{\tau}_n(\rho)X_nl_1\nn\\
 &&+(\mu_4-3\sigma^4)
 \{l_2^2||Vec(diag\tilde{G_n})||^2+2l_2l_3tr(\tilde{G_n})+nl_3^2\}\nn\\
 &&+2\mu_3\{l_1^{\tau} X^{\tau}_nA_n(\rho)Vec(diag \tilde{G}_n)l_2+l_1^{\tau}X^{\tau}_nA_n(\rho) {\mathbf{1}}_n l_3\}\nn\\
&=& l^\tau \Sigma_{k+2}l,\nn
\eea
where $\Sigma_{k+2}$ is given in (\ref{Sigma}).
From Condition A3, one can see that $n^{-1}\sigma^2_Q\geq c_1>0$. From Lemma \ref{lemm4.2}, we have
\[
{Q_n-E(Q_n)\over \sigma_Q }\tod N(0, 1).
\]
Noting that $E(Q)=0$, we thus have (\ref{22}).

Next we will prove (\ref{23}),  i. e.
\bea
n^{-1}\sum_{i=1}^n (l^\tau\omega_i(\theta))^2=n^{-1}\sigma_Q^2+o_p(1).\label{wn2}
\eea
Let
\bea
Y_{in}&=& l^\tau\omega_i(\theta)\nn\\
&=&u_{ii}(\epsilon^2_i-\sigma^2)+2\sum^{i-1}_{j=1}u_{ij}\epsilon_i\epsilon_j+v_{i}\epsilon_i\nn\\
&=&u_{ii}(\epsilon^2_i-\sigma^2)+B_i\epsilon_i,
\eea
where $B_i=2\sum^{i-1}_{j=1}u_{ij}\epsilon_j+v_{i}$. Let  ${\mathcal{F}}_{0}=\{ {\emptyset}, \Omega\}, {\mathcal{F}}_{i}=\sigma(\epsilon_1, \epsilon_2, \cdots, \epsilon_i), 1\leq i\leq n$. Then $\{Y_{in}, {\mathcal{F}}_{i}, 1\leq i\leq n\}$ form a martingale difference array.
Note that
\bea
&& n^{-1}\sum_{i=1}^n \{ l^\tau\omega_i(\theta) \}^2-n^{-1}\sigma_Q^2 = n^{-1}\sum_{i=1}^n(Y_{in}^2-EY_{in}^2)\nn\\
&=& n^{-1}\sum_{i=1}^n\{Y_{in}^2-E(Y_{in}^2|{\mathcal{F}}_{i-1})+E(Y_{in}^2|{\mathcal{F}}_{i-1})-EY_{in}^2 \}\nn\\
&=& n^{-1}S_{n1}+n^{-1}S_{n2},
\eea
where $S_{n1}=\sum_{i=1}^n\{Y_{in}^2-E(Y_{in}^2|{\mathcal{F}}_{i-1})\}$, $S_{n2}=\sum_{i=1}^n, \{E(Y_{in}^2|{\mathcal{F}}_{i-1})-EY_{in}^2\}$.
Next we will prove
\bea
n^{-1}S_{n1}=o_p(1),\label{sn1}
\eea
and
\bea
n^{-1}S_{n2}=o_p(1).\label{sn2}
\eea
It suffices to prove $n^{-2}E(S^2_{n1})\to 0$ and
 $n^{-2}E(S_{n2})^2\to 0$ respectively. Obviously,
 \[
 Y_{in}^2=u_{ii}^2(\epsilon^2_i-\sigma^2)^2+B_i^2\epsilon_i^2+2u_{ii}B_i(\epsilon^2_i-\sigma^2)\epsilon_i.
 \]
Thus
\[
E(Y_{in}^2|{\mathcal{F}}_{i-1})=u_{ii}^2E(\epsilon^2_i-\sigma^2)^2+B_i^2\sigma^2+2u_{ii}B_i\mu_3.
\]
It follows that
\bea
&&n^{-2}E(S^2_{n1})=n^{-2}\sum_{i=1}^nE\{Y_{in}^2-E(Y_{in}^2|{\mathcal{F}}_{i-1})\}^2\nn\\
&=&n^{-2}\sum_{i=1}^n E[u_{ii}^2\{(\epsilon^2_i-\sigma^2)^2-E(\epsilon^2_i-\sigma^2)^2\}+B_i^2(\epsilon_i^2-\sigma^2)\nn\\
&& +2u_{ii}B_i
(\epsilon^3_i-\sigma^2\epsilon_i-\mu_3)]^2\nn\\
&\leq& Cn^{-2}\sum_{i=1}^n E[u_{ii}^4\{(\epsilon^2_i-\sigma^2)^2-E(\epsilon^2_i-\sigma^2)^2\}^2]+Cn^{-2}\sum_{i=1}^n E\{B_i^4(\epsilon_i^2-\sigma^2)^2\}\nn\\
&&+Cn^{-2}\sum_{i=1}^n E\{u_{ii}^2B_i^2(\epsilon^3_i-\sigma^2\epsilon_i-\mu_3)^2\}.
\label{sn11}
\eea
By Condition A1, we have
\bea
n^{-2}\sum_{i=1}^n E[u_{ii}^4\{(\epsilon^2_i-\sigma^2)^2-E(\epsilon^2_i-\sigma^2)^2\}^2]\leq C n^{-1}\to 0,\label{sn12}
\eea
and
\bea
&&n^{-2}\sum_{i=1}^n E\{B_i^4(\epsilon_i^2-\sigma^2)^2\}\leq C n^{-2}\sum_{i=1}^n E(\sum^{i-1}_{j=1}u_{ij}\epsilon_j+v_{i})^4\nn\\
&\leq& C n^{-2}\sum_{i=1}^n E(\sum^{i-1}_{j=1}u_{ij}\epsilon_j)^4+ C n^{-2}\sum_{i=1}^nv_{i}^4\nn\\
&\leq& C n^{-2}\sum_{i=1}^n \sum^{i-1}_{j=1}u_{ij}^4\mu_4+C n^{-2}\sum_{i=1}^n (\sum^{i-1}_{j=1}u_{ij}^2\sigma^2)^2+C n^{-2}\sum_{i=1}^n (l_1^{\tau}b_i)^4\nn\\
&\leq& C n^{-1}\to 0.\label{sn13}
\eea
Similarly, one can show that
\be
n^{-2}\sum_{i=1}^n E\{u_{ii}^2B_i^2(\epsilon^3_i-\sigma^2\epsilon_i-\mu_3)^2\} \to 0. \label{sn13.2}
\ee
From (\ref{sn11})-(\ref{sn13.2}), we have $n^{-2}E(S_{n1}^2)\to 0$.
Furthermore,
\bea
E(Y_{in}^2)&=&E\{E(Y_{in}^2|{\mathcal{F}}_{i-1})\}=u_{ii}^2E(\epsilon^2_i-\sigma^2)^2+\sigma^2E(B_i^2)+2u_{ii}\mu_3E(B_i)\nn\\
&=& u_{ii}^2E(\epsilon^2_i-\sigma^2)^2+\sigma^2(4\sum^{i-1}_{j=1}u_{ij}^2\sigma^2+v_i^2)+2u_{ii}\mu_3v_i.\nn
\eea
Thus,
\bea
&& n^{-2}E(S_{n2}^2)=n^{-2}E[\sum_{i=1}^n\{E(Y_{in}^2|{\mathcal{F}}_{i-1})-EY_{in}^2\}]^2\nn\\
&=& n^{-2}E[\sum_{i=1}^n \{B_i^2\sigma^2-\sigma^2(4\sum^{i-1}_{j=1}u_{ij}^2\sigma^2+v_i^2)+2u_{ii}\mu_3(B_i-v_i)\}]^2\nn\\
&=& n^{-2}\sum_{i=1}^n E[\sigma^2\{(2\sum^{i-1}_{j=1}u_{ij}\epsilon_j)^2-4\sum^{i-1}_{j=1}u_{ij}^2\sigma^2\}
+4(\sum^{i-1}_{j=1}u_{ij}\epsilon_j)v_i\sigma^2\nn\\
&& +2u_{ii}\mu_3(2\sum^{i-1}_{j=1}u_{ij}\epsilon_j)]^2\nn\\
&\leq& Cn^{-2}\sum_{i=1}^nE\{\sigma^2(\sum^{i-1}_{j=1}u_{ij}\epsilon_j)^2-\sum^{i-1}_{j=1}u_{ij}^2\sigma^2\}^2
+Cn^{-2}\sum_{i=1}^nE\{(\sum^{i-1}_{j=1}u_{ij}\epsilon_j)v_i\sigma^2\}^2\nn\\
&&+Cn^{-2}\sum_{i=1}^nE\{2u_{ii}\mu_3(\sum^{i-1}_{j=1}u_{ij}\epsilon_j)\}^2.\label{sn21}
\eea
Note that
\bea
&&n^{-2}\sum_{i=1}^nE[\sigma^2\{(\sum^{i-1}_{j=1}u_{ij}\epsilon_j)^2-\sum^{i-1}_{j=1}u_{ij}^2\sigma^2\}]^2\leq n^{-2}\sigma^4\sum_{i=1}^n
E(\sum^{i-1}_{j=1}u_{ij}\epsilon_j)^4\nn\\
&\leq & Cn^{-2}\sum_{i=1}^n\sum^{i-1}_{j=1}u_{ij}^4\mu_4+C n^{-2}\sum_{i=1}^n (\sum^{i-1}_{j=1}u_{ij}^2\sigma^2)^2
\leq C n^{-1}\to 0,\label{sn22}
\eea
\bea
n^{-2}\sum_{i=1}^nE\{(\sum^{i-1}_{j=1}u_{ij}\epsilon_j)v_i\sigma^2\}^2=n^{-2}\sigma^6\sum_{i=1}^nv_i^2\sum^{i-1}_{j=1}u_{ij}^2
\leq C n^{-2}\to 0,\label{sn23}
\eea
and
\bea
n^{-2}\sum_{i=1}^nE\{2u_{ii}\mu_3(\sum^{i-1}_{j=1}u_{ij}\epsilon_j)\}^2=4\mu_3^2\sigma^2n^{-2}\sum_{i=1}^nu_{ii}^2\sum^{i-1}_{j=1}u_{ij}^2
\leq C n^{-1}\to 0,\label{sn24}
\eea
where we have used Conditions A1 and A2.
From (\ref{sn21})-(\ref{sn24}), we have $n^{-2}ES_{n2}^2\to 0$.  The proof of  (\ref{wn2}) is thus complete.

Finally, we will prove (\ref{24}). Note that
\bea
\sum_{i=1}^n E||\omega_i(\theta)||^3 &\leq& \sum_{i=1}^nE||b_i\epsilon_i||^3+\sum_{i=1}^nE|\tilde{g}_{ii}(\epsilon^2_i-\sigma^2)
+2\epsilon_i\sum^{i-1}_{j=1}\tilde{g}_{ij}\epsilon_j|^3\nn\\
&&+\sum_{i=1}^nE|\epsilon^2_i-\sigma^2|^3.\label{31}
\eea
By Conditions A1 and A2,
\bea
\sum_{i=1}^nE||b_i\epsilon_i||^3 \leq Cn(\max_{1\leq i \leq n}||x_i||)^3E|\epsilon_1|^3=O(n),\label{32}
\eea
\bea
&&\sum_{i=1}^nE \left |\tilde{g}_{ii}(\epsilon^2_i-\sigma^2)
+2\epsilon_i\sum^{i-1}_{j=1}\tilde{g}_{ij}\epsilon_j \right |^3 \nn\\
&\leq & C\sum_{i=1}^nE|\tilde{g}_{ii}(\epsilon^2_i-\sigma^2)|^3+
C\sum_{i=1}^nE \left |2\epsilon_i\sum^{i-1}_{j=1}\tilde{g}_{ij}\epsilon_j \right |^3
\nn\\
&\leq & C\sum_{i=1}^nE|\tilde{g}_{ii}(\epsilon^2_i-\sigma^2)|^3+
C\sum_{i=1}^nE|\epsilon_i|^3\sum^{i-1}_{j=1}E|\tilde{g}_{ij}\epsilon_j|^3\nn\\
&&¡¡+
C\sum_{i=1}^nE|\epsilon_i|^3¡¡\left \{ \sum^{i-1}_{j=1}E(\tilde{g}_{ij}\epsilon_j)^2 \right \}^{3/2}
=O(n),\label{33}
\eea
\bea
\sum_{i=1}^nE|\epsilon^2_i-\sigma^2|^3=O(n).\label{34}
\eea
From (\ref{31})-(\ref{34}),we have
\bea
\sum_{i=1}^n E||\omega_i(\theta)||^3=O(n).\label{w3}
\eea
Further, using (\ref{w3}) and Markov inequality, we obtain $\sum_{i=1}^n ||\omega_i(\theta)||^3=O_p(n)$. Thus (\ref{24})
is proved.

We now in the position to prove the main results in this article.

{\bf  Proof of Theorem \ref{theo2.1}. } Let $\lambda=\lambda(\theta), \rho_0=||\lambda||,
\lambda=\rho_0\eta_0$. From (\ref{9}), we have
\[
\frac{\eta_0^{\tau}}{n}\sum_{j=1}^{n}\omega_{j}(\theta)-\frac{\rho_0}{n}\sum_{j=1}^{n}
{(\eta_0^{\tau}\omega_{j}(\theta))^2\over
1+\lambda^{\tau} \omega_{j}(\theta)}=0.
\]
It follows that
\[
|\eta_0^{\tau}\bar{\omega}|\geq
{\rho_0\over 1+\rho_0
Z_{n}}\lambda_{min}(S_0),
\]
where $Z_{n}$ is defined in (\ref{21}), $\bar{\omega}=n^{-1}\sum^n_{i=1}\omega_i(\theta), S_0=n^{-1}\sum^n_{i=1}\omega_i(\theta)\omega^{\tau}_i(\theta)$.
That is
\[
| \eta_0^{\tau}\Sigma^{1/2}_{k+2}\Sigma^{-1/2}_{k+2}\bar{\omega}|\geq
{\rho_0\over 1+\rho_0
Z_{n}}\lambda_{min}(S_0),
\]
i. e.
\[
\lambda_{max}(\Sigma^{1/2}_{k+2})|| \eta_0||\cdot ||\Sigma^{-1/2}_{k+2}\bar{\omega}||\geq
{\rho_0\over 1+\rho_0
Z_{n}}\lambda_{min}(S_0).
\]
Combining with Lemma \ref{lemm4.4} and Condition A3, we have
\[
{\rho_0\over 1+\rho_0  Z_{n}}=O_p(n^{-1/2}).
\]
Therefore, from Lemma \ref{lemm4.4},
\[
\rho_0=O_p(n^{-1/2}).
\]
Let
$\gamma_i=\lambda^{\tau}\omega_{i}(\theta)$. Then \be \max_{1\leq i \leq
n}|\gamma_i|=o_p(1).  \label{gam}
\ee
Using (\ref{9})  again, we have
\begin{eqnarray*}
0 & = & {1\over {n}}\sum_{j=1}^{n}{\omega_{j}(\theta)\over 1+\lambda^{\tau} \omega_{j}(\theta)}\\
& = & {1\over {n}}\sum_{j=1}^{n} \omega_{j}(\theta)-{1\over
{n}}\sum_{j=1}^{n}{\omega_{j}(\theta)\{\lambda^{\tau}\omega_{j}(\theta)\}
\over 1+\lambda^{\tau} \omega_{j}(\theta)}\\
& = & {1\over {n}}\sum_{j=1}^{n} \omega_{j}(\theta)- \{{1\over
{n}}\sum_{j=1}^{n}\omega_{j}(\theta)\omega_{j}(\theta)^{\tau}\}\lambda+
{1\over {n}}\sum_{j=1}^{n}{\omega_{j}(\theta)\{\lambda^{\tau}\omega_{j}(\theta)\}^2\over 1
+\lambda^{\tau} \omega_{j}(\theta)}\\
& = & {1\over {n}}\sum_{j=1}^{n} \omega_{j}(\theta)-  \{{1\over
{n}}\sum_{j=1}^{n}\omega_{j}(\theta)\omega_{j}(\theta)^{\tau}\}\lambda+ {1\over {n} }
\sum_{j=1}^{n}\frac{\omega_{j}(\theta)\gamma_j^2}{1+\gamma_j}\\
&=& \overline{\omega}- S_0\lambda+{1\over {n} }
\sum_{j=1}^{n}\frac{\omega_{j}(\theta)\gamma_j^2}{1+\gamma_j}.
\end{eqnarray*}
Combining with Lemma \ref{lemm4.4} and Condition A3,  we may write
\be
\lambda=S_0^{-1}\overline{\omega}+\varsigma,  \label{lambda000}
\ee
 where  $||\varsigma||$ is  bounded by
\[
 n^{-1}\sum_{j=1}^{n}||\omega_{j}(\theta)||^3||\lambda||^2
=O_p(n^{-1}).
\]
By (\ref{gam}) we may
expand $\log(1+\gamma_i)=\gamma_i-\gamma_i^2/2+\nu_i$  where, for
some finite $B>0$,
$$
P(|\nu_i|\leq B|\gamma_i|^3, 1\leq i \leq n)\rightarrow 1, \hbox{ as } n\rightarrow \infty.
$$
Therefore, from (\ref{8}), (\ref{lambda000}) and Taylor expansion, we have
\begin{eqnarray*}
\ell_n(\theta)& = & 2\sum_{j=1}^{n} \log(
1+\gamma_j)=2\sum_{j=1}^{n}
\gamma_j-\sum_{j=1}^{n}\gamma_j^2+2\sum_{j=1}^{n}\nu_j\\
& = & 2n\lambda^{\tau}\overline{\omega}-n\lambda^{\tau}S_0\lambda+2\sum_{j=1}^{n}\nu_j\\
&=&2n(S_0^{-1}\overline{\omega})^{\tau}\overline{\omega}+2n\varsigma^{\tau}\overline{\omega}
-n\overline{\omega}^{\tau}S_0^{-1}\overline{\omega}-\\
&&2n\varsigma^{\tau}\overline{\omega}
-n\varsigma^{\tau}S_0\varsigma+2\sum_{j=1}^{n}\nu_j\\
&=&n\overline{\omega}^{\tau}S_0^{-1}\overline{\omega}-
n\varsigma^{\tau}S_0\varsigma+2\sum_{j=1}^{n}\nu_j\nn\\
&=&  \{n\Sigma^{-1/2}_{k+2}\overline{\omega}\}^{\tau}
\{n\Sigma^{-1/2}_{k+2} S_0 \Sigma^{-1/2}_{k+2}\}^{-1}
\{n\Sigma^{-1/2}_{k+2}\overline{\omega}\}\nn\\
&& -n\varsigma^{\tau}S_0\varsigma+2\sum_{j=1}^{n}\nu_j.\nn
\end{eqnarray*}
From Lemma \ref{lemm4.4} and Condition A3, we have
\[
 \{n\Sigma^{-1/2}_{k+2}\overline{\omega}\}^{\tau}  \{n\Sigma^{-1/2}_{k+2} S_0 \Sigma^{-1/2}_{k+2})\}^{-1}
\{n\Sigma^{-1/2}_{k+2}\overline{\omega}\} \tod \chi^2_{k+2}.
\]
On the other hand, using Lemma \ref{lemm4.4} and above derivations, we can see that  $n\varsigma^{\tau}S_0\varsigma=O_p(n^{-1})=o_p(1)$
and
\[
|\sum_{j=1}^{n}\nu_j|\leq B ||\lambda||^3\sum_{j=1}^{n}||\omega_{j}(\theta)||^3 =O_p(n^{-1/2})=o_p(1).
\]
 The proof of Theorem \ref{theo2.1} is thus complete.


\bigskip
\noindent {\bf Acknowledgements  }
 This work was partially
supported by the National Natural Science Foundation of China
(11671102), the Natural Science Foundation of Guangxi (2016GXNSFAA3800163, 2017GXNSFAA198349) and the Program on the High Level  Innovation Team and Outstanding
Scholars in Universities of Guangxi Province.

\bigskip
\noindent {\bf References}

\noindent

\nh Anselin, L., 1988, Spatial Econometrics: Methods and Models. The Netherlands: Kluwer Academic
Publishers.

\nh Anselin, L. and Bera, A. K., 1998,  Spatial dependence in linear regression models with
an introduction to spatial econometrics, Handbook of Applied Economics Statistics, ed. by
Ullah A. and Giles, D. E. A.,  New York: Marcel Dekker.

\nh Bell K. P. and  Bockstael, N. E., 2000, Applying the generalized-moments estimation approach
to spatial problems involving microlevel data, The Review of Economics and Statistics,
82, 72-82.

\nh Besley, T. and  Case, A., 1995,  Incumbent behavior: vote-seeking, tax-Setting, and yardstick
competition, The American Economic Review, 85, 25-45.

\nh Bertrandm, M, Luttmer, E. F. P. and Mullainathan, S., 2000, Network effects and welfare
cultures, Quarterly Journal of Economics, 115, 1019-1055.

\nh Brueckner, J. K., 1998, Testing for strategic interaction among local governments: the
case of growth controls, Journal of Urban Economics, 44, 438-467.

\nh Case, A. C., 1991, Spatial patterns in household demand, Econometrica, 59, 953-965.

\nh Case, A. C., Rosen, H. S. and Hines, J. R., 1993, Budget Spillovers and fiscal policy interdependence:
evidence from the States, Journal of Public Economics, 52, 285-307.

\nh Chen, J., Qin, J., 1993,  Empirical likelihood estimation for finite populations and the effective usage of auxiliary information,
Biometrika, 80, 107-116.

\nh Chen, S. X. and Keilegom, I. V. 2009, A review on empirical likelihood for regressions (with discussions), Test, 3, 415-447.

\nh Cliff, A. D. and Ord, J. K., 1973, Spatial Autocorrelation. London: Pion Ltd.

\nh Cressien, N., 1993, Statistics for Spatial Data, New York: John Wiley \& Sons.

\nh Dow, M. M., Burton, M. L., and White, D. R., 1982, Network autocorrelation:
a simulation study of a foundational problem in
regression and survey research, Social Networks, 4, 169-200.

\nh Kelejian, H. H.  and  Prucha, I. R., 1999, A generalized moments estimator for the autoregressive parameter in a spatial
model, International Economic Review, 40, 509-533.

\nh Kelejian, H. H., Prucha, I. R., 2001, On the asymptotic distribution of the Moran $I$ test statistic with applications, Journal of Econometrics, 104, 219-257.

\nh  Kr\"{a}mer, W. and  Donninger, C., 1987,  Spatial autocorrelation among errors and the
relative efficiency of OLS in the linear regression model, Publications of the American Statistical Association, 82, 577-579.

\nh Lee, L. F., 2004, Asymptotic distributions of quasi-maximum likelihood estimators for spatial autoregressive models, Econometrica, 72, 1899-1925.

\nh Ord, K. 1979, Estimation methods for models of spatial interaction,¡±
Journal of the American Statistical Association, 70, 120-126.

\nh  Owen, A. B., 1988,  Empirical likelihood ratio confidence intervals for a single functional,
 Biometrika, 75, 237-249.

\nh  Owen, A. B., 1990, Empirical likelihood ratio confidence regions,
Ann. Statist., 18, 90-120.

\nh   Owen, A. B., 1991, Empirical likelihood for linear models,  Ann. Statist., 19, 1725-1747.

\nh Owen, A. B., 2001, Empirical Likelihood, London: Chapman \& Hall.

\nh  Qin, J. and Lawless, J., 1994, Empirical likelihood and general estimating equations,
Ann. Statist., 22, 300-325.

\nh Topa, G. 2001, Social interactions, local Spillovers and unemployment, Review of Economic
Studies, 68, 261-295.

\nh Wu, C. B., 2004, Weighted empirical likelihood inference, Statistics \& Probability Letters,  66, 67-79.

\nh Zhong, B.,  Rao, J.N.K., 2000, Empirical likelihood inference under stratified random sampling using auxiliary population
information,  Biometrika, 87, 929-938.

\newpage

\begin{table}
\centering

  \caption{Coverage probabilities of the LR and EL confidence regions with $\epsilon_i\sim N(0,1)$}\label{1}
  \vspace{.1in}

\begin{tabular}{cccc|cccc}

\hline
$\rho$ &$W_n$ & LR & EL &$\rho$  &$W_n$ &LR & EL\\
\hline
 -0.85&$grid_{49}$ &0.9715&0.8760&-0.15 & $grid_{49}$ &0.9435&0.8820\\
      & $grid_{100}$ &0.9655&0.9200&& $grid_{100}$ &0.9450 &0.9045\\
      &$grid_{169}$& 0.9595&0.9370& &$grid_{169}$ &0.9455&0.9325\\
      & $W_{49}$ &0.9630&0.8840& & $W_{49}$ &0.9405& 0.8645\\
      & $I_5\bigotimes W_{49}$ &0.9565& 0.9370& & $I_5\bigotimes W_{49}$ &0.9455&0.9330\\
      & $W_{345}$ &0.9535 &0.9260& &   $W_{345}$&0.9460&0.9395\\
 \hline
 0.85&$grid_{49}$ &0.9285&0.8635&0.15 & $grid_{49}$ &0.9290&0.8680\\
      & $grid_{100}$ &0.9320&0.9045&& $grid_{100}$ & 0.9435&0.9160\\
      &$grid_{169}$& 0.9435&0.9305& &$grid_{169}$ &0.9470&0.9320\\
      & $W_{49}$ &0.9435&0.8680& & $W_{49}$ &0.9450& 0.8805\\
      & $I_5\bigotimes W_{49}$ &0.9560& 0.9500& & $I_5\bigotimes W_{49}$ &0.9525&0.9405\\
      & $W_{345}$ &.9545 &0.9445& &   $W_{345}$&0.9485&0.9375\\
\hline
\end{tabular}
\end{table}

\begin{table}
\centering

  \caption{Coverage probabilities of the LR and EL confidence regions with $\epsilon_i\sim t(5)$}\label{1}
  \vspace{.1in}

\begin{tabular}{cccc|cccc}

\hline
$\rho$ &$W_n$ & LR & EL &$\rho$  &$W_n$ &LR & EL\\
\hline
  -0.85&$grid_{49}$ &0.8640&0.8025&-0.15 & $grid_{49}$ &0.8695&0.8010\\
      & $grid_{100}$ &0.8575&0.8610&& $grid_{100}$ &0.8310 &0.8640\\
      &$grid_{169}$& 0.8400&0.8870& &$grid_{169}$ &0.8160&0.8800\\
      & $W_{49}$ &0.8670&0.8065& & $W_{49}$ &0.8355&0.7990 \\
      & $I_5\bigotimes W_{49}$ &0.8425&0.9155 & & $I_5\bigotimes W_{49}$ &0.8175&0.8930\\
      & $W_{345}$ &0.8145 &0.9010& &   $W_{345}$&0.8290&0.9200\\
 \hline
 0.85&$grid_{49}$ &0.8180&0.7890&0.15 & $grid_{49}$ &0.8520&0.8040\\
      & $grid_{100}$ &0.8160&0.8575&& $grid_{100}$ &0.8440 &0.8750\\
      &$grid_{169}$& 0.8115&0.9020& &$grid_{169}$ &0.8210&0.8970\\
      & $W_{49}$ &0.8480&0.7855& & $W_{49}$ &0.8495&0.7985 \\
      & $I_5\bigotimes W_{49}$ &0.8180& 0.9010& & $I_5\bigotimes W_{49}$ &0.8090&0.8955\\
      & $W_{345}$ &0.8030&0.9110& &   $W_{345}$&0.8065&0.9125\\
\hline

\end{tabular}
\end{table}

\begin{table}
\centering

  \caption{Coverage probabilities of the LR and EL confidence regions with $\epsilon_i+4\sim \chi^2_4$}\label{1}
  \vspace{.1in}

\begin{tabular}{cccc|cccc}

\hline
$\rho$ &$W_n$ & LR & EL &$\rho$  &$W_n$ &LR & EL\\
\hline
  -0.85&$grid_{49}$ &0.8670&0.8070&-0.15 & $grid_{49}$ &0.8560&0.8080\\
      & $grid_{100}$ &0.8530&0.8850&& $grid_{100}$ &0.8370 &0.8610\\
      &$grid_{169}$&0.8570 &0.8950& &$grid_{169}$ &0.8450&0.8975\\
      & $W_{49}$ &0.8615&0.7985& & $W_{49}$ &0.8490&0.8125 \\
      & $I_5\bigotimes W_{49}$ &0.8580& 0.9185& & $I_5\bigotimes W_{49}$ &0.8385&0.9160\\
      & $W_{345}$ &0.8525 &0.9270& &   $W_{345}$&0.8275&0.9295\\
 \hline
 0.85&$grid_{49}$ &0.8365&0.7915&0.15 & $grid_{49}$ &0.8505&0.7955\\
      & $grid_{100}$ &0.8320&0.8530&& $grid_{100}$ &0.8430 &0.8690\\
      &$grid_{169}$& 0.8395&0.8900& &$grid_{169}$ &0.8320&0.9050\\
      & $W_{49}$ &0.8490&0.7820& & $W_{49}$ &0.8445&0.7920 \\
      & $I_5\bigotimes W_{49}$ &0.8435&0.9050 & & $I_5\bigotimes W_{49}$ &0.8385&0.9215\\
      & $W_{345}$ &0.8490 &0.9325& &   $W_{345}$&0.8430&0.9285\\
\hline

\end{tabular}
\end{table}

\end{document}